\documentstyle[12pt,epsf]{article}
\newlength{\oline}
\newlength{\tline}
\begin{document}
\setlength{\baselineskip}{\tline}
\setlength{\parskip}{\oline}
\parindent 0.0cm 
\setcounter{page}{1}

\begin{center}
{\bf \Large 
Hybrid Quantum and Classical Mechanical Monte Carlo Simulations of the 
Interaction of Hydrogen Chloride with Solid Water Clusters }
\end{center}

\underline {Dar\' \i o A. Estrin} $^{a,b}$,
\underline {Jorge Kohanoff} $^{b}$,
Daniel H. Laria $^{a,c}$,
and Ruben O. Weht $^{b,c}$ 

\begin{center}
{\it
$^a$ Departamento de Qu\' \i mica Inorg\'anica, Anal\' \i tica y 
Qu\' \i mica-F\' \i sica e INQUIMAE,\\
Facultad de Ciencias Exactas y 
Naturales, Universidad de Buenos Aires\\
Ciudad Universitaria, Pabell\'on II, 1428, Buenos Aires, ARGENTINA.\\
$^b$ International Centre for Theoretical Physics\\
Strada Costiera 11, 34014, Trieste, ITALY.\\
$^c$ Comisi\'on Nacional de Energ\' \i a At\'omica, \\
Avenida Libertador 8250, 1429, Buenos Aires, ARGENTINA.}
\end{center}

\par Monte Carlo simulations using a hybrid quantum and classical mechanical
potential were performed for crystal and amorphous-like
HCl(H$_{2}$O)$_{n}$ clusters (n $\le$ 24). The subsystem composed by
HCl and one water molecule was treated within Density Functional Theory,
and a classical force field was used for the rest of the system.
Simulations performed at 200~$K$ suggest that the energetic feasibility of
HCl dissociation strongly depends on  its initial placement
within the cluster.
An important degree
of ionization occurs only if HCl is incorporated
into the surface. We observe that local melting does not play a
crucial role in the ionization process.

\newpage

\section{Introduction}

\par Atmospheric chemistry is a research area in which many relevant 
processes occur in heterogeneous environments, such as the surface of 
solid particles and within liquid droplets. In particular, investigations
connected with the stratospheric ozone layer have proved that ionic
solvation of HCl at the surface of ice crystals is an important source 
of chlorine atoms, which may ultimately induce ozone-destroying chain 
reactions~\cite{abbat,molina,hanson}.

Simulations of HCl dissociation at ice surfaces using classical
force fields have recently been reported~\cite{hynes,clary1}. These
are based on parametrizations of the potential energy surface which are
derived from gas phase calculations for the isolated HCl(H$_{2}$O) dimer
and the ionic complex Cl$^-$ + H$_3$O$^+$. Situations like those
described above, in which a chemical reaction is strongly influenced 
by the environment, are rather delicate, and a purely classical approach
risks of exhibiting problems of potentials transferability.
A quantum mechanical semiempirical study has also been reported for HCl 
solvated in water clusters~\cite{buesnel}. This calculation does take 
into account in a better way the effects of the environment, but it shows 
a very poor performance for the isolated HCl-acceptor water subsystem. 
This is a consequence of the limitations of the semiempirical description
of the quantum mechanical Hamiltonian. Full {\it ab initio} Car-Parrinello 
simulations of HCl dissociation have been performed, although in a bulk 
water environment~\cite{kari}. 

In order to use an accurate electronic structure technique and be able to
sample adequately configuration space at an affordable computational cost, 
we have devised a hybrid approach~\cite{ho,estrin} in which the HCl-acceptor 
water subsystem is treated at the Density Functional Theory (DFT) 
level~\cite{metecc} and the rest of the system is modeled using the 
TIP4P potential for water~\cite{klein}. We have also analyzed the role of the 
initial conditions and the local melting on the energetic feasibility of 
HCl dissociation in crystal-like and amorphous-like clusters 
HCl(H$_{2}$O)$_{n}$ ($n  \le $ 24) at a temperature of 200 K using Monte 
Carlo simulation techniques.

\section{The hybrid QM/CM strategy }

\par The computational scheme is constructed by partitioning the system into
a quantum mechanical (QM) and a classical mechanical (CM) region~\cite{ho}.
Considering $N_c$ atoms in the classical subsystem with coordinates and
partial charges  \{${\bf R}_i, q_i, i=1,\cdots,N_c$\} 
and $N_q$ atoms in the QM region with coordinates and nuclear charges 
\{${\bf \tau}_{\alpha}, z_{\alpha},~\alpha=1,\cdots,N_q$\}
the total energy can be written as:
\begin{equation}
E[\rho]=E_{\rm KS}[\rho]~+~\sum_{i=1}^{Nc}{q_i}\int\frac{\rho({\bf r})}
{\mid{\bf r}-{\bf R}_i\mid}~{\rm d}{\bf r}~+~
\sum_{i=1}^{Nc}\sum_{\alpha=1}^{Nq} [ v_{\rm LJ}(\mid{\bf R}_i-{\bf\tau}_
{\alpha}\mid)~+
\frac {q_i z_{\alpha}} {\mid{{\bf R}_{i} - {\bf \tau}_{\alpha}} \mid}  ]
+~E_{\rm CM}~.
\end{equation}

\par In this equation the first term is a purely quantum mechanical piece 
given by the standard Kohn-Sham expression~\cite{ks}. The electronic 
density $\rho$ is obtained by solving a Kohn-Sham set of equations 
self-consistently, where the external potential contribution to the 
Kohn-Sham operator includes the electrostatic interaction with the
CM region, as given by the second and third terms of expression (1).
The second term accounts for the electrostatic interaction of the
charges representing the atoms (or molecules) situated in the CM region
with the electronic charge distribution, while the third term corresponds to 
the Van der Waals and electrostatic interactions between the nuclei in
the CM region and those in the QM region. TIP4P parameters~\cite{klein}
were used for O and H, and Lennard-Jones parameters for Cl were taken 
from~\cite{cloro}. The last term, $E_{CM}$, is the classical solvent 
contribution, and has been modeled with a flexible TIP4P potential for 
water which includes harmonic stretching and bending intramolecular terms 
extracted from extensive {\it ab initio} calculations~\cite{bartlett}. 
The electrostatic interactions between nuclei in the QM region are included 
in the Kohn-Sham expression (first term).

For the QM region, computations are performed at the generalized 
gradient approximation (GGA) level. The correlation part is composed by 
the parametrization of the homogeneous electron gas due to Vosko~\cite{vwn}
and the gradient corrections given by Perdew~\cite{perdew}. The local 
exchange term was supplemented with the gradient corrections proposed
by Becke~\cite{becke1}. The exchange-correlation contribution to the 
potential and the energy is calculated by a numerical integration scheme 
based on grids and quadratures also proposed by Becke~\cite{becke2}. 
Gaussian basis sets are used for the expansion of the one-electron orbitals
and also for the additional auxiliary set used for expanding the electronic
density. Double zeta plus polarization basis sets have been employed for
Cl, O and H~\cite{godbout}. Auxiliary sets were also taken from 
Ref.~\cite{godbout}.

In order to check the accuracy of the QM part of the Hamiltonian, geometry
optimizations and vibrational analysis have been performed for isolated 
HCl, H$_{2}$O and HCl(H$_{2}$O). Structural results are shown in Table 1, 
together with results recently obtained at the MP2 level~\cite{clary2}. 
The agreement between DFT-GGA and MP2 results as well as with available
experimental data is rather satisfactory. This is consistent with previous 
work~\cite{sim,clementi} in which DFT calculations at the GGA level
proved to perform well for hydrogen-bonded dimers.

The performance of the QM/CM approach was tested by computing the binding 
energies, structural parameters and vibrational frequencies of the 
clusters HCl(H$_{2}$O)$_{n}$ (n=2,3), considering the subsystem formed by 
HCl and the acceptor water molecule as the QM subsystem, and the remaining 
water molecules as the CM part. For these clusters, MP2 and also some 
experimental results are available. Selected structural parameters are shown 
in Table 1. The agreement with the MP2 computed values is again reasonable. 
An increase of the HCl and a decrease of the OCl bond length with cluster 
size can be observed, implying that the H-bond strength increases with 
the number of water molecules in the cluster.

Binding energies and the $\nu_{HCl}$ vibrational stretching frequencies 
for HCl(H$_{2}$O)$_{n}$ (n=2,3) are reported in Table 2, compared with 
experimental data and MP2 results. A red shift 
in the HCl stretching frequency is experimentally observed upon complexation 
with water molecules, and reproduced by theoretical calculations. This 
implies that proton transfer is increasingly favored in larger clusters.
Interaction energies for the HCl(H$_{2}$O) complex show a good agreement
with MP2 calculations. Results for  larger clusters 
show an overestimation of binding energies seemingly because of the use of
a TIP4P classical potential parametrized for bulk water. However, the 
errors are expected to become less important for larger aggregates, as  
one approaches the bulk situation.

\vspace*{1.5cm}

\section{Monte Carlo simulations}

\par Finite temperature properties were simulated using a Monte Carlo (MC) 
technique~\cite{charu}. The MC moves consisted of random changes in the
positions of all the particles simultaneously (including intramolecular
solvent motions), with maximum displacements independent of their respective 
masses. The standard Metropolis sampling algorithm was used~\cite{metropolis}, 
and the maximum displacements were adjusted to give an overall acceptance 
ratio of about 50\%. Ensemble averages were calculated over 
15000 trial moves in all cases, after 4000 moves of equilibration. All
simulations were carried out at 200 $K$, a temperature which is 
characteristic of stratospheric conditions.

We have considered the following situations:

\begin{enumerate}
\item{ HCl(H$_{2}$O), hereafter referred as Case~1.}
\item{ HCl(H$_{2}$O)$_{16}$ amorphous-like clusters. The initial 
conditions for these clusters have been obtained by running classical 
MC simulations at 200 K, in which the structure of the HCl(H$_{2}$O) 
dimer was constrained during the course of the simulation. This was 
achieved by fixing the HCl bond length to 1.34~\AA (Case 2A) and 
1.90~\AA (Case 2B), respectively. In case 2A, the HCl molecule remained 
at the periphery of the cluster, while in case 2B it was 
incorporated into the surface.}

\item{ HCl(H$_{2}$O)$_{24}$ crystal-like clusters. The initial 
conditions were generated by isolating a fragment of two bilayers of 
hexagonal ice, composed by 25 water molecules, and replacing an
appropriately oriented water molecule with an HCl. In the first case
we replaced a water molecule situated in the outer layer (case 3A). 
In the second and third cases, the water molecule replaced was selected
in the second layer (cases 3B and 3C). In case 3B, the orientation
of the HCl molecule was chosen such that it was H-bonded to a water 
acceptor molecule located in the first layer, while in case 3C the HCl 
was H-bonded to a water molecule located in the second layer.}
These are typical configurations that are likely to be found
during the ice growth process under stratospheric
conditions~\cite{hynes}.
\end{enumerate}

Schematic views for cases 1, 2A, and 2B are shown in Figure 1, and for
cases 3A, 3B, and 3C in Figure 2. 

Radial distribution functions $g(r)$ for H-Cl, H-O(acceptor water) and 
Cl-O(acceptor water) are presented in Figure 3 for cases 1, 2A, and 2B 
and in Figure 4 for cases 3A, 3B and 3C. It can be observed that an 
important extent of ionization occurs in cases 2B, 3B and 3C, $i.e.$ in
those situations where the HCl is incorporated into the
surface, instead of remaining as an adsorbate. The degree of ionization, 
however, is not complete. This can be seen in the first peak of $g(r)$ 
for H-O(acceptor water), which lies at about 1.2 -- 1.3~\AA, while the 
optimized HO bond distance in [H$_{3}$O]$^{+}$ is about 1.0~\AA.

It can be observed in Figure 3 that no ionization occurs in case 2A, 
where the HCl peak in the $g(r)$ remains at about the equilibrium distance
of the isolated HCl molecule. The different behavior observed in cases 
2A and 2B can be explained in terms of the solvation of the products, which 
is determined basically by the initial conditions. [H$_{3}$O]$^{+}$
prefers trigonal coordination, and situations in which it acts as an 
acceptor in H-bonds are unfavorable. On the other hand, Cl$^{-}$
prefers maximum H coordination. In case 2B, [H$_{3}$O]$^{+}$ would be
trigonally coordinated as well as the chloride ion, but in case 2A the 
chlorine is found in the periphery of the cluster and solvation is rather 
poor.

Figure 4 shows that there is no dissociation in case 3A. This is because
Cl results with only coordination 2 and [H$_{3}$O]$^{+}$ would act as an 
acceptor of an H bond (tetrahedral coordination). In both 3B and 3C cases 
dissociation occurs. The larger degree of ionization observed in Case 3B 
is due to the fact that, while Cl always exhibits a trigonal coordination,
in case 3C the [H$_{3}$O]$^{+}$ is tetrahedrally solvated, and in 3B it
has the optimal trigonal coordination. It is also interesting to remark
the different behavior observed for the $g(r)$ for O-Cl 
in the different simulations. 
In the case in which HCl is in the outer layer (case 3A), it peaks at 
about 3.00~\AA ~and in cases in which it is in the second monolayer 
(3B and 3C), it peaks at 2.76~\AA ~and 2.77~\AA, respectively. The same 
trend is observed in the amorphous-like clusters (2A and 2B), for which 
$g(r)$ peaks at a larger value when HCl is not dissociated.

Ensemble averages for Cl Mulliken population, H-Cl, H-O(acceptor water), 
Cl-O(acceptor water) distances and binding energies are given in Table 3. 
More negative averages for the Cl Mulliken populations are consistent with 
the large degree of dissociation observed in cases 2B, 3B and 3C. It can 
also be noted that larger average binding energies per molecule are 
associated with the better solvated (larger extent of ionization) 
situations. In all simulations the clusters remained solid-like, at least 
in the region of phase space sampled during our simulations. Values of 
Lindemann's relative rms bond length fluctuations were typically 0.02. 
Melting phenomena have not been observed, even in the simulations with an 
important degree of dissociation.

\vspace*{1.5cm}

\section{Conclusions}

\par We conclude that the energetic feasibility for HCl ionization in 
solid-like clusters strongly depends on the initial placement of the HCl 
within the system, which in turn determines the solvation properties of the 
products. Local melting phenomena turn out not to be necessarily related to 
the dissociation process. Our results on crystal-like clusters reinforce 
the conclusions of Ref.~\cite{hynes}, in which simulations of HCl 
incorporated into bulk-ice surfaces were performed using classical 
potentials. Moreover, we have shown that the same conclusion holds for
amorphous-like clusters. In the case of HCl adsorbed on top of ice 
surfaces it appears that the HCl dissociation process would not be 
energetically favorable~\cite{clary1}. These observations also show 
that the accuracy of the Hamiltonian description plays a fundamental role 
in these studies. The QM/CM Monte Carlo scheme proposed in this work 
provides an accurate tool for modeling chemical reactions in 
heterogeneous environments. 

Before closing this article, we would like to make
a final comment concerning ergodicity and proper sampling of
all relevant fluctuations. During our MC runs the systems remained 
well-equilibrated and we did not observe any signature of 
transitions between the
different solvation structures described in the previous paragraphs. This
clearly shows the presence of a high free energy 
barrier - in comparison to normal thermal energies - implying 
that, in principle, 
the feasibility of the dissociation process would  
be strongly dependent on the initial solvation conditions, $i.e.$ on the 
details of the growth process. 
In any event, one would tend to believe that the more energetically
favorable configuration, namely the one with
the larger negative solvation energy (2B or 3B in our studies, see Table~3) 
would correspond to the most stable configuration from the thermodynamic
point of view. However to be certain,
a more complete analysis involving the computation of relative free energies
between the different solvation structures is necessary; this would allow us
to estimate not only equilibrium information 
but also information about rates of interconvertion between different solvation
structures. Work in this direction is currently being undertaken.

\vspace*{1.5cm}

\section*{Acknowledgments}

\par D.A.E. acknowledges Fundaci\'on Antorchas and Universidad de Buenos Aires 
for financial support and ICTP for hospitality. We thank also Francesco 
Sciortino for providing us with ice configurations and Roberto Fern\'andez 
Prini for bringing this problem to our attention and for useful discussions.

\newpage

\begin{table}
{\bf TABLE 1: Selected optimized geometrical parameters for 
HCl, H$_{2}$O and HCl(H$_{2}$O)$_{n}$ (n=1,3)
with bond lengths in~\AA~and angles in deg. $<$OHCl
is the hydrogen bond angle and
dO$\cdots$Cl and  dO$\cdots$H the hydrogen bond lengths.}\\

\vspace*{1.0cm}

\begin{center}
\begin{tabular}{llcccc}
\hline
  & &  &  &  & \\
  & &  DFT $^a$   &  MP2 $^b$    &  MP2 $^c$   & Exp.  \\
  &  & & & & \\
\hline
 & & & & & \\
 HCl   &    d HCl  & 1.286  & 1.271   & 1.281   & 1.275 $^d$ \\
 & & & & & \\
 H$_{2}$O  & d HO  & 0.981  & 0.961 & 0.968 & 0.958 $^d$ \\
           & $<$HOH  & 104.8 & 103.5  & 104.8 & 104.5 $^d$ \\
 & & & & & \\
 HCl(H$_{2}$O)      & d HCl           & 1.320 & 1.287 & 1.302 & \\
                   & d O $\cdots$ Cl & 3.095 & 3.196 & 3.120 & 3.2149 $^e$ \\
                   & d O $\cdots$ H  & 1.776 & 1.910 & 1.818 & \\
                   & $<$OHCl        & 176.6 & 176.7 & 178.7 & \\
 & & & & & \\
 HCl(H$_{2}$O)$_2$ & d HCl           & 1.343 & 1.303 & 1.326 & \\
                   & d O $\cdots$ Cl & 2.992 & 3.059 & 2.993 & \\
                   & d O $\cdots$ H  & 1.672 & 1.787 & 1.688 & \\
                   & $<$OHCl        & 165.6 & 163.3 & 166.5 & \\
 & & & & & \\
 HCl(H$_{2}$O)$_3$ & d HCl           & 1.369 & 1.323 &  & \\
                   & d O $\cdots$ Cl & 2.923 & 2.976 & & \\
                   & d O $\cdots$ H  & 1.558 & 1.657 & & \\
                   & $<$OHCl        & 174.2 & 174.7 & & \\
 & & & & & \\
\hline
\end{tabular}
\end{center}
$^a$ this work.\\
$^b$ 6-31g(2dp) results of Ref.~\cite{clary2}.\\
$^c$ Pol1 results of Ref.~\cite{clary2}.\\
$^d$ Ref.~\cite{huber}.\\
$^e$ Ref.~\cite{legon}.\\
\end{table}

\newpage

\begin{table}
{\bf TABLE 2:  Binding energies (kJ/mol) and $\nu_{HCl}$ stretching frequencies
(cm$^{-1}$) for HCl and HCl(H$_{2}$O)$_{n}$. (n=1,3) $^a$ } 

\vspace*{1.0cm}

\begin{center}
\begin{tabular}{llcccc}
\hline
  & &  &  &  & \\
  & &  DFT $^b$   &  MP2 $^c$    &  MP2 $^d$   & Exp.  \\
  &  & & & & \\
\hline
 & & & & & \\
 HCl   &    $\nu_{HCl}$  & 2967   & 3068   & 2982   & 2991  $^e$ \\
 & & & & & \\
 HCl(H$_{2}$O)      & $\nu_{HCl}$     & 2512 & 2841 & 2709  & 2659 $^f$ \\
                   &               &      &      &       &  2540 $^g$ \\
                   & $\Delta$E$_e$ & 23.71  & 21.97  & 20.57 & \\
                   & $\Delta$E$_o$ & 14.18  & 13.74  & 10.93  & \\
 & & & & & \\
 HCl(H$_{2}$O)$_2$ & $\nu_{HCl}$     & 2257   & 2615 & 2394 & 2390$^f$ \\
                   & $\Delta$E$_e$ & 67.79  & 51.27 & 50.95  & \\
                   & $\Delta$E$_o$ & 50.04  & 30.53 & 32.86  & \\
 & & & & & \\
 HCl(H$_{2}$O)$_3$ & $\nu_{HCl}$     & 2015  & 2341  &  &  \\
                   & $\Delta$E$_e$ & 118.04 & 89.07  &  & \\
                   & $\Delta$E$_o$ & 89.00  & 54.70  &  & \\
 & & & & & \\
\hline
\end{tabular}
\end{center}
$^a$  $\Delta$E$_e$ 
is the cluster dissociation  energy,
$\Delta$E$_o$ includes also
zero point energy corrections.\\
$^b$ this work.\\
$^c$ 6-31g(2dp) results of Ref.~\cite{clary2}.\\
$^d$ Pol1 results of Ref.~\cite{clary2}.\\
$^e$ Ref.~\cite{huber}.\\
$^f$ experimental results in Ar matrix (Ref.~\cite{amirand}).\\
$^g$ experimental results in N$_{2}$ matrix (Ref.~\cite{ault}).\\
\vspace*{1cm}
\end{table}

\newpage

\begin{table}
{\bf TABLE 3:  Ensemble averages of Cl Mulliken population, H-Cl, 
H-O(acceptor water),
Cl-O(acceptor water) bond distances (\AA), and binding energy
per molecule (kJ/mol). Values in parenthesis are standard deviations.}

\vspace*{1.0cm}

\begin{center}
\begin{tabular}{lccccc}
\hline
 & & & & & \\
    & qCl  & d H-Cl  & d H-O  & Cl-O  & E  \\
 & & & & & \\
\hline
  & &  &  &  & \\
 1  & -0.208 (0.017) & 1.314 (0.026)  & 1.914 (0.118) &  3.185 (0.107)& -7.7 (1.5) \\
  & &  &  &  & \\
 2A  & -0.245 (0.017) & 1.325 (0.027) & 1.752 (0.066) & 3.061 (0.057) & -34.0 (0.6) \\
  & &  &  &  & \\
 2B  & -0.528 (0.036) & 1.496 (0.047) & 1.298 (0.052) & 2.789 (0.039) & -36.6 (0.7) \\
  & &  &  &  & \\
 3A  & -0.280 (0.017) & 1.344 (0.030) & 1.672 (0.068) & 2.997 (0.068) & -34.4 (0.4) \\
  & &  &  &  & \\
 3B  & -0.546 (0.027) & 1.510 (0.040)& 1.257 (0.037) & 2.760 (0.043) & -35.9 (0.5) \\
  & &  &  &  & \\
 3C  & -0.464 (0.032) & 1.440 (0.040)& 1.335 (0.049) & 2.769 (0.043) & -34.6 (0.6) \\
  & &  &  &  & \\
\hline
\end{tabular}
\vspace*{7cm}
\end{center}
\end{table}

\newpage

{\Large\bf Figure Captions}\\

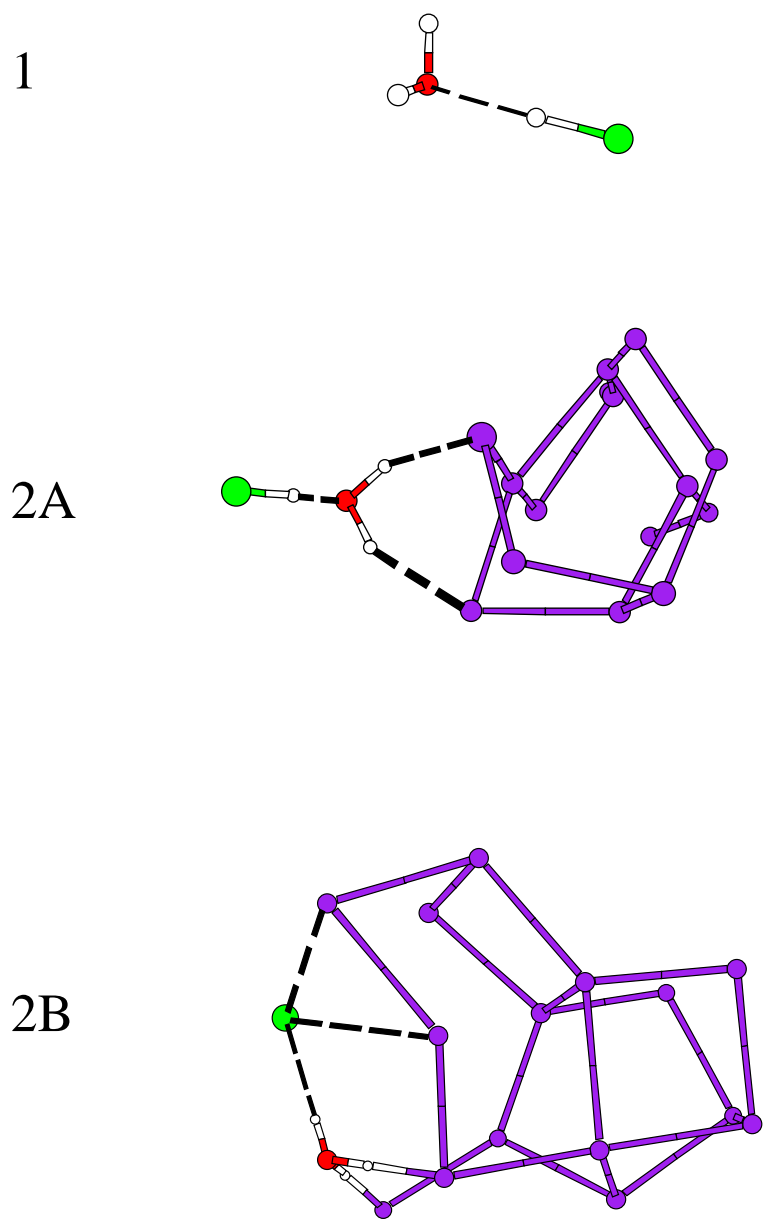
Figure 1:\\
Schematic view of initial conditions.
HCl(H$_{2}$O) ( Case 1)  and HCl(H$_{2}$O)$_{15}$  (Cases 2A and 2B).
Only H in the QM subsystem are shown. Relevant H bonds are represented
with dashed lines.\\

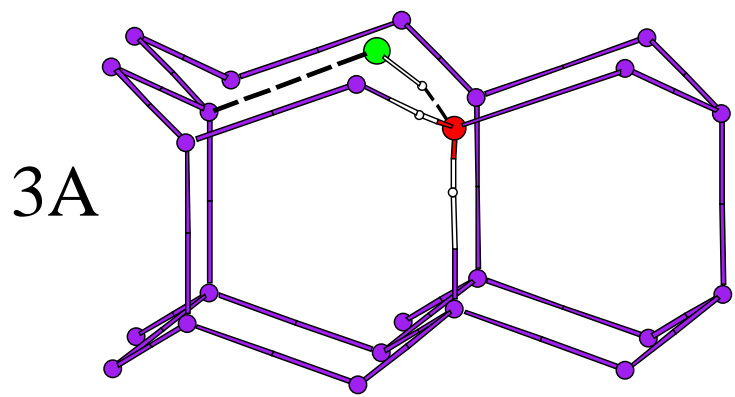
Figure 2:\\
Schematic view of initial conditions.
HCl(H$_{2}$O)$_{24}$ (Cases 3A, 3B and 3C).
Only H in the QM subsystem are shown. Relevant H bonds are represented
with dashed lines.\\
 
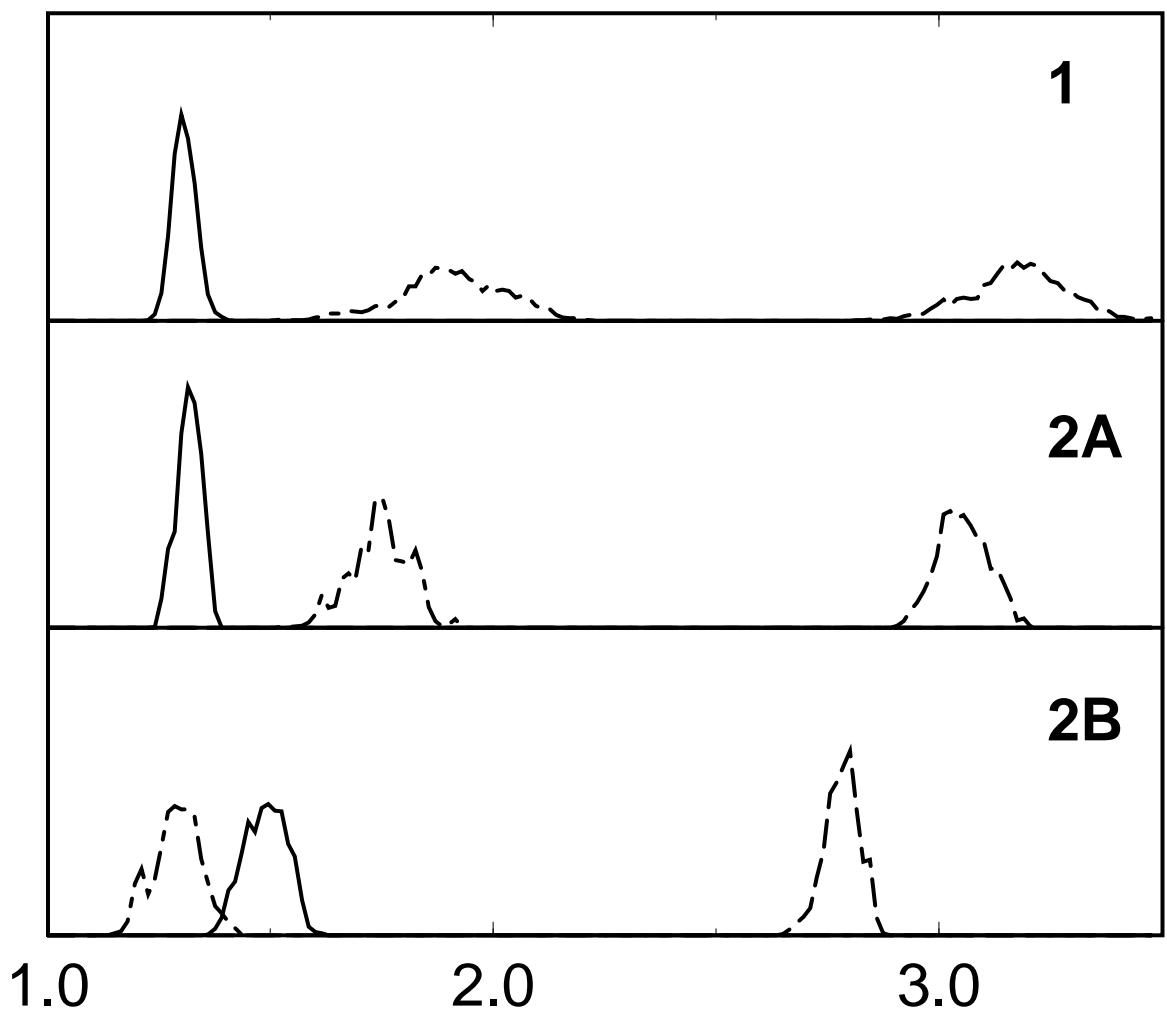
Figure 3:\\
H-Cl (solid line), H-O(acceptor water) (dashed-dotted line)
and Cl-O(acceptor water) (dashed line) radial correlation functions, for
cases 1, 2A and 2B. (distances in $\AA$)\\
 
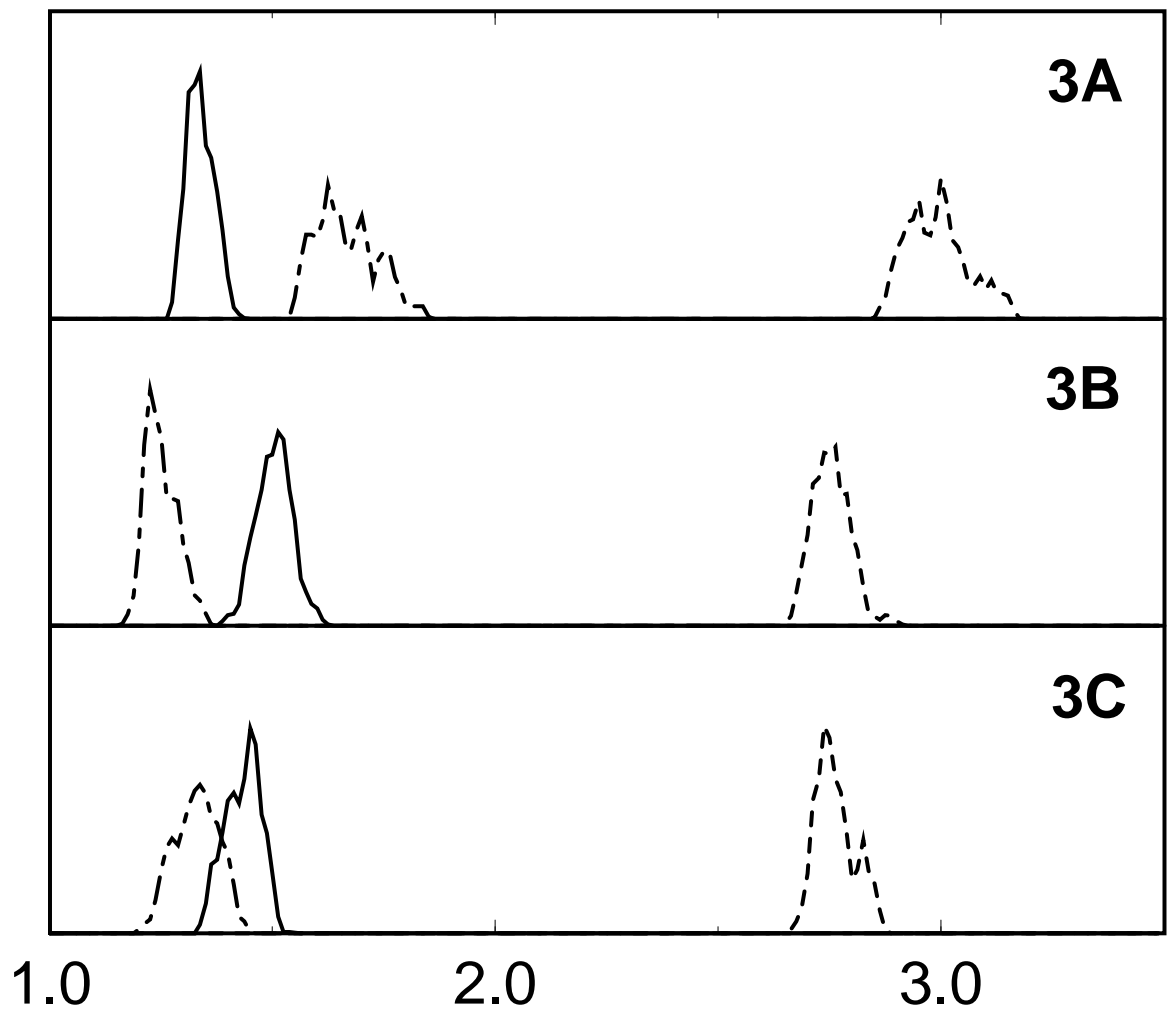
Figure 4:\\
H-Cl (solid line), H-O(acceptor water) (dashed-dotted line)
and Cl-O(acceptor water) (dashed line) radial correlation functions, for
cases 3A, 3B and 3C. (distances in $\AA$)
 

\newpage

\begin{thebibliography}{99}

\bibitem{abbat} J.P.D. Abbat and M.J. Molina, J. Phys. Chem. 96 (1992) 7674.
\bibitem{molina} M.J. Molina, T.L. Tso, L.T. Molina, and E.Y. Yang, Science
238 (1987) 1253. 
\bibitem{hanson} D.R. Hanson and A.R. Ravishankara, J. Phys. Chem. 96 (1992)
2682.
\bibitem{hynes} B.J. Gertner and J.T. Hynes, Science 271 (1996) 1563.
\bibitem{clary1} S.H. Robertson and D.C. Clary, J. Chem. Soc. Faraday
Discussions, 100 (1995) 309.
\bibitem{buesnel} R. Buesnel, I.H. Hillier, and A.J. Masters, Chem. Phys. Lett.
247 (1995) 391.
\bibitem{kari} K. Laasonen and M.L. Klein, J. Am. Chem. Soc. 116 (1994) 11620.
\bibitem{ho} L.L. Ho, A.D. MacKerell Jr., and P.A. Bash, J. Phys.
Chem. 100 (1996) 4466.
\bibitem{estrin} D.A. Estrin, L. Liu, and S.J. Singer, J. Phys. Chem. 96 (1992)
5325.
\bibitem{metecc} D.A. Estrin, G. Corongiu, and E. Clementi, in: METECC, 
Methods and Techniques in Computational Chemistry, ed. Clementi, E. (Stef, 
Cagliari, 1993) chapter 12.
\bibitem{klein} W.L. Jorgensen, J. Chandrasekar, J.D. Madura, R.W. Impey, and
M.L. Klein, J. Chem. Phys. 79 (1983) 926.
\bibitem{ks} W. Kohn and L.J. Sham, Phys. Rev. A 140 (1965) 1133.
\bibitem{cloro} M.P. Allen and D.J. Tildesley, Computer simulations of
liquids (Clarendon Press, Oxford, 1987).
\bibitem{bartlett} R.J. Bartlett, I. Shavitt, and G.D. Purvis, J. Chem. 
Phys. 71 (1979) 281.
\bibitem{vwn} S.H. Vosko, L. Wilk, and  M. Nusair, Can. J. Phys. 58 
(1980) 1200.
\bibitem{perdew} J.P. Perdew, Phys. Rev. B, 33 (1986) 8822;
Erratum, Phys. Rev. B 34 (1986) 7406.
\bibitem{becke1} A.D. Becke, Phys. Rev. A 38 (1988) 3098.
\bibitem{becke2} A.D. Becke, J. Chem. Phys. 88 (1988) 1053.
\bibitem{godbout} N. Godbout, D. R. Salahub, J. Andzelm, and E. Wimmer, 
Can. J. Chem. 70 (1992) 560.
\bibitem{clary2} M.J. Packer and D.C. Clary, J. Phys. Chem. 99 
(1995) 14323.
\bibitem{sim} F. Sim, A. St-Amant, I. Papai, and D.R. Salahub, J. Am. Chem. 
Soc.  114 (1992) 4391.
\bibitem{clementi} D.A. Estrin, L. Paglieri, G. Corongiu, and E. Clementi,
J. Phys. Chem. 100 (1996) 8701.
\bibitem{huber} K.P. Huber and G. Herzberg, Molecular Spectra and
Molecular Structure (Van Nostrand Reinhold, New York, 1979), Vol. IV.
\bibitem{legon} A.C. Legon and L.C. Willoughby, Chem. Phys. Lett. 95 (1983) 
449.
\bibitem{amirand} C. Amirand and D. Maillard, J. Mol. Struct. 176 (1988) 181.
\bibitem{ault} B.S. Ault and G.C. Pimentel, J. Phys. Chem. 77 (1973) 57.
\bibitem{charu} R.O. Weht, J. Kohanoff, D. A. Estrin, and C. Chakravarty
(submitted).
\bibitem{metropolis} N. Metropolis, A.W. Rosenbluth, M.N. Rosenbluth, 
A.H. Teller and E. Teller, J. Chem. Phys. 21 (1953) 1087.
\end{thebibliography}
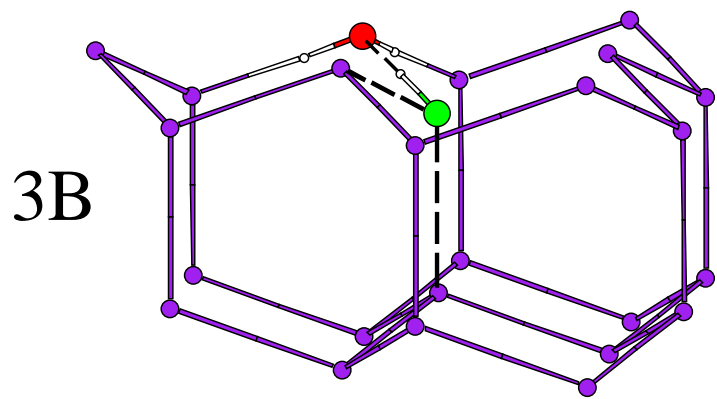
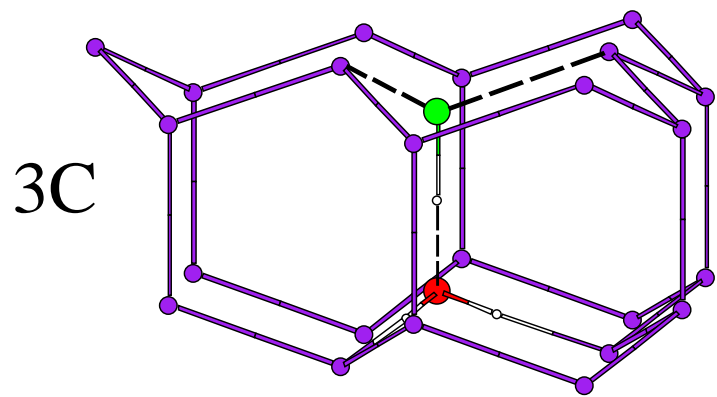
\end{document}